# Design and Validation of a MATLAB-based GUI for Coarray Domain Analysis of Sparse Linear Arrays

Ashish Patwari[1*], Ananya Pandey[2], Aditya Dabade[2], Priyadarshini Raiguru[3]

[1]Assistant Professor, School of Electronics Engineering, Vellore Institute of Technology, Vellore, Tamil Nadu, India

[2]Outgoing Student, School of Electronics Engineering, Vellore Institute of Technology, Vellore, Tamil Nadu, India

[3]Assistant Professor, Department of ECE, ITER, SOA deemed to be university, Bhubaneswar, Odisha, India

*Corresponding author ORCID ID - https://orcid.org/0000-0001-9489-7004

Email IDs: ashish.p@vit.ac.in; ananya.pandey2021@vitstudent.ac.in; aditya.dabade2021@vitstudent.ac.in; priyadarshiniraiguru@soa.ac.in

*Abstract*—This work presents a first-of-its-kind graphical user interface (GUI)-based simulator developed using MATLAB App designer for the comprehensive analysis of sparse linear arrays (SLAs) in the difference coarray (DCA) domain. Sparse sensor arrays have emerged as a critical solution in enhancing signal detection, direction of arrival (DOA) estimation, and beamforming in fields such as wireless communication, radar, sonar, and integrated sensing systems. They offer several advantages over traditional uniform arrays, including reduced system complexity, lower deployment costs, and improved mitigation of mutual coupling effects. The tool enables users to input array configurations, compute DCAs, visualize weight function graphs, and assess the hole-free status of arrays, as applicable for coarray processing. Unlike conventional simulators that focus on radiation pattern visualization (array pattern, main lobe and sidelobe characteristics, azimuth cut, rectangular view, polar view etc.), this tool addresses the behavior of SLAs from a coarray domain perspective. Numerical validations demonstrate the tool's correctness, effectiveness, and its potential to foster further research in sparse arrays. This simulator could also be used as a teaching aid to drive home complicated topics and attract young minds towards the fascinating field of sparse array design.

## I. Introduction

Sensor arrays have been widely studied in various fields such as seismology, radar, sonar, biomedical imaging, wireless communication, and radio astronomy [1], [2], [3]. Arrays have high sensitivity and can detect faint signals that may go unnoticed if there is a single sensor. An array can also sense directions, unlike individual antennas, and allows for electronic beam steering (by adjusting the phase angles of individual elements) as opposed to mechanical beam steering. An active array has the additional ability to adjust its response according to the changing signal environment [1]. Antenna arrays can perform two major functions: direction of arrival (DOA) estimation and/or beamforming. A smart antenna array performs both these tasks [4].

Sparse arrays require fewer sensors than uniform arrays do to provide the same aperture. As a result, they aid in reducing system costs, power consumption, and deployment/maintenance costs. Sparse arrays operate with little or no degradation in performance compared with uniform arrays in terms of DOA estimation. They can detect more sources than the available number of sensors and are less affected by mutual coupling than uniform arrays [5], [6], [7].

While minimum redundancy and minimum hole arrays have been known for almost six decades, the past decade witnessed unprecedented developments in the design of sparse array structures for various purposes. Several novel arrays, including the coprime array, the two-level nested array, the MISC array, and their variants were introduced in the past 15 years [5], [8], [9], [10], [11], [12], [13], [14], [15], [16], [17], [18]. A comprehensive review of sparse linear arrays and their properties can be found in [19] and [20]. Invaluable insights into sensor arrays and their applications can be found in [21], [22], [23].

*Categories of sparse arrays*: Sparse linear arrays (SLAs) prove to be advantageous over ULAs as the former lend themselves to difference coarray processing. Most of the aforementioned arrays are designed in such a way that their difference coarray is hole-free. This aids in unambiguous DOA estimation. While continuous difference coarray is a requirement for passive DOA estimation, active DOA estimation requires a continuous sum coarray [24], [25]. Similarly, SLAs with continuous sum and difference co-array (SDCA) are required to estimate the directions of non-circular sources [26], [27]. Sparse MIMO radar arrays, on the other hand, play a major role in active DOA estimation in automotive radar applications (mostly in advanced driver assistance systems (ADAS) to steer clear of the limitations that exist in the delay-doppler domain) [28], [29], [30]. Nevertheless, most modern SLAs have been designed with a focus on difference coarray processing.

*Importance of sparse arrays in future technologies*: Recent research has shown that sparse arrays have a prominent role to play in next generation wireless technologies such as integrated sensing and communications (ISAC), physical layer security, and location-based services. A large number of engineers and researchers well-versed in sparse arrays, DOA estimation, and beamforming will be required in the future to tackle the workforce demands of future technologies. Additionally, the importance of SLAs in radar, sonar, and biomedical fields cannot be undermined. In this context, there is a need to develop simple and interactive tools that can be used by non-specialist and novice users. This can reduce the entry barrier and improve the uptake of the field. The designed tool shall provide a quick summary of the characteristics possessed by the array under test (AUT).

*A. Functions of the proposed simulator*

In this work, we present a GUI developed using MATLAB App designer, that incorporates all essential metrics for sparse array analysis in the difference coarray (DCA) domain. The GUI design allows users to input the AUT, either in terms of its sensor positions or IES notation. The simulator computes the DCA, weight function, primary weights, and determines the hole-free status of the AUT, as shown in Fig. 1.

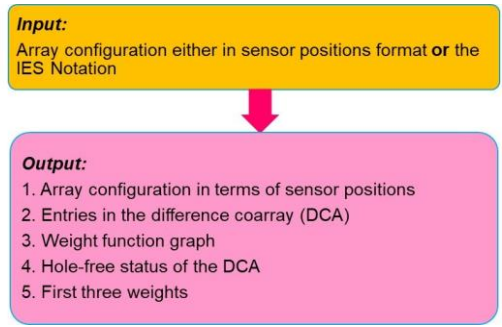

Fig. 1. Input and output parameters of the proposed GUI

*B. Motivation and Contributions*

Although many array simulators are available in the existing literature [31], [32], [33], [34], [35],[36] their main functionality is to visualize the array radiation pattern, main lobe width, sidelobe levels etc., which are useful for array synthesis. *To the best of our knowledge, there are no simulators that study the coarray behavior of sparse linear arrays.* Hence, we designed the proposed GUI and made it available for public usage and evaluation. The underlying MATLAB codes that run in the back end are also shared. We believe that this tool would cut short the learning curve and encourage more people to get involved in sparse array research. The specific contributions of this paper are:

  i. We designed the first-ever GUI that helps in visualizing the coarray properties of SLAs. The GUI was designed from scratch by meticulously planning its intended features and the layout of various components.

  ii. The functionality of the tool was validated by cross-checking the properties of well-known SLAs available in the current literature. In particular, it was ensured that the tool makes no mistakes when computing the DCA entries, location of holes, hole-free status, and weight function.

The rest of the paper is organized as follows. Section II explains the mathematical framework used for sparse array analysis. Section III explains the methodology used to design the GUI simulator. Section IV presents numerical examples that validate the functionality and correctness of the simulator. Section V outlines the limitations and future scope of the designed GUI. Section VI concludes the paper.

## II. Sparse Array Fundamentals

This section discusses important terminology related to sparse arrays and coarray processing required to understand the upcoming sections.

## A. Coarray fundamentals

### 1) The Difference Set and Spatial Lags

Set $\mathbb{S}$ denotes the positions of the sensors in the physical array, normalized to half the wavelength. All attainable self and cross spacings between the sensor positions in $\mathbb{S}$ form the difference set $\mathbb{Z}$. Each entry in $\mathbb{Z}$ is called a spatial/coarray lag.

### 2) Difference Coarray and Holes

The non-repeating and sorted entries of $\mathbb{Z}$ form the difference coarray (DCA), denoted by $\mathbb{D}$. Missing spatial lags in $\mathbb{Z}$ and hence in $\mathbb{D}$ are known as coarray holes and are undesirable.

### 3) The Weight Function

The weight $w(m)$ denotes the number of times a spatial lag $m$ appears in the difference set $\mathbb{Z}$. The weight function lists out the weights of all spatial lags present in the DCA.

### 4) Inter-element spacing (IES) notation

The IES notation lists the spacing between consecutive array elements and has been widely used in the past to represent sparse arrays. An array with IES notation $\{a, b, c, d\}$ has physical sensors at $\{0, a, a + b, a + b + c, a + b + c + d\}$, relative to the half wavelength.

## B. Signal Model

The signal model is based on second order coarray domain processing. The coarray correlation matrix is formed and subjected to eigen value decomposition in order to form the signal and noise subspaces. The same is widely available in existing literature [5], [15], [16], and is hence not repeated here.

## III. GUI Design Methodology

The methodology for designing the proposed MATLAB GUI is explained below. At first, the characteristics and features to be incorporated into the GUI have been outlined. Because the main function of the simulator is to aid in the coarray analysis of the AUT, the following functions were deemed necessary. Fig. 2 shows the step-by-step breakdown of the functionalities expected from the proposed GUI.

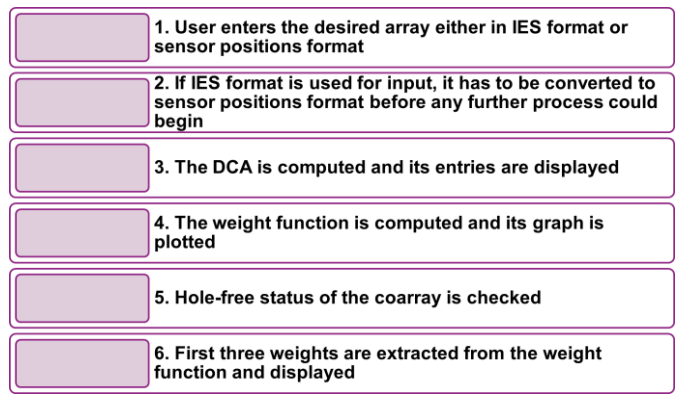

Fig. 2. Overview of the GUI's desired functionality

*A. Planning the user interface (UI)*

The MATLAB app designer allows users to create professional grade apps without requiring the knowledge of advanced software development skills. It is based on simple drag and drop of visual components and uses the integrated editor to efficiently program it's functionality.

The component library's drop-down feature was employed to select whether the AUT should be defined using IES notation or sensor positions. Once an option is selected, the input is captured into a string variable. The UI axes component was utilized to visualize the weight function. A text label was used to convey the hole-free status of the AUT. Other output parameters such as sensor positions, DCA entries, and primary weights are displayed using a text area. A push button was used to clear previous memory and to process the new input array. Following the arrangement of these components in the design view, the code view of the app designer was accessed to incorporate the callback function for the push button, which delineates the application's functionality. Upon successful implementation, the application will be ready for execution.

*B. Code segments to realize different functions*

The following code segments were used to enable the different components of the GUI.

- For all the analysis, it is essential to know the positions of the physical sensors in the array. The entries in the IES notation (if selected by the user) were converted to actual sensor positions using the program mentioned in Appendix A.1.
- Once the sensor positions are obtained, the difference set and difference coarray were computed using the code given in Appendix A.2.
- Following this, the weight function was computed and plotted using the commands given in Appendix A.3.
- The hole-free status of the array was verified using the commands mentioned in Appendix A.4.
- The primary (first three) weights were extracted and displayed using the command mentioned in Appendix A.5. Only three weights are displayed because it is widely accepted in sparse array theory that these weights influence the array's susceptibility to mutual coupling.

The complete GUI was then bundled as a standalone app for installation and download into the MATLAB app corner.

## IV. VALIDATING THE GUI SIMULATOR

On opening the GUI, the following home screen appears as shown in Fig. 3. All labels and placeholders for input and output parameters can be seen clearly. The tool is available for download and use from https://github.com/ananyapandey-gui/SparseArrayAnalyser .

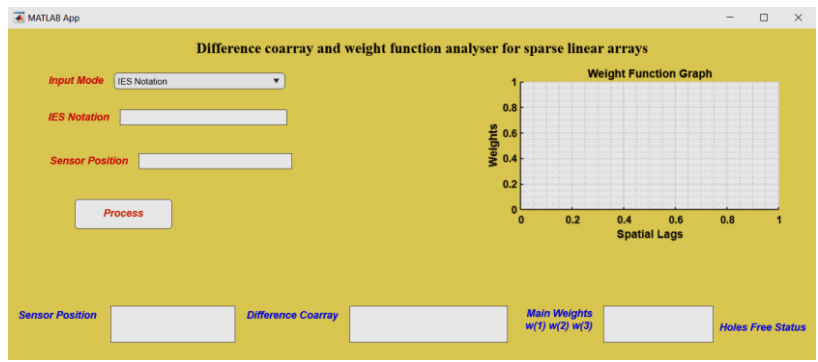

Fig. 3. Home screen of the simulator

We now verify whether the simulator produces accurate outputs as desired by using famous sparse arrays from the existing literature.

*A. Small and simple arrays*

To keep things simple at the beginning, we started out with small arrays whose properties could be verified by hand calculations. Firstly, the array with sensor positions [0, 1, 4, 6] was considered. This array qualifies as a zero redundancy MRA and as a perfect MHA as well. Coarray analysis of this array is shown in Fig. 4. As expected, the DCA is continuous from -6 to +6, and the weight function has no discontinuities. Next, the coarray analysis was performed on the array [0, 1, 2, 6] and the result is shown in Fig. 5. The DCA listings and the weight function graph confirm the presence of a hole at $\pm 3$.

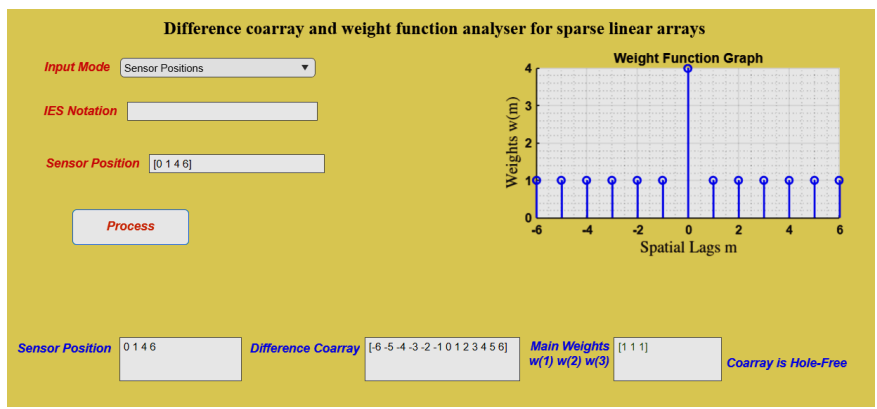

Fig. 4. Coarray analysis of [0, 1, 4, 6]

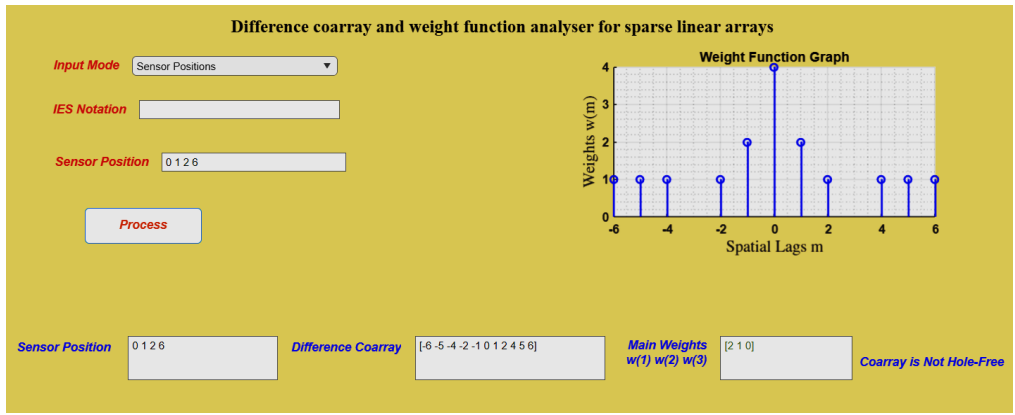

Fig. 5. Coarray analysis of [0, 1, 2, 6]

B. *Arrays with coarray discontinuities*

It is known that coprime arrays, optimally dense non redundant arrays (ODNRAs), and weight-constrained sparse arrays (WCSAs) contain holes in their respective DCAs. Hence, it was imperative to test these arrays to verify the simulator's correctness. The 6-element coprime array [0, 2, 3, 4, 6, 9], the 6-element ODNRA [0, 4, 6, 7, 15, 20], and the 10-element WCSA were considered for analysis. Fig. 6 shows the results for the coprime array and Fig. 7 shows the analysis for the ODNRA. It can be seen from Fig. 6 that the coprime array has holes at $\pm 8$ while Fig. 7 reveals that the ODNRA has many holes. The weight function of the ODNRA matches the one given in part-a of Fig. 1. in Ebrahimi and Karimi [37].

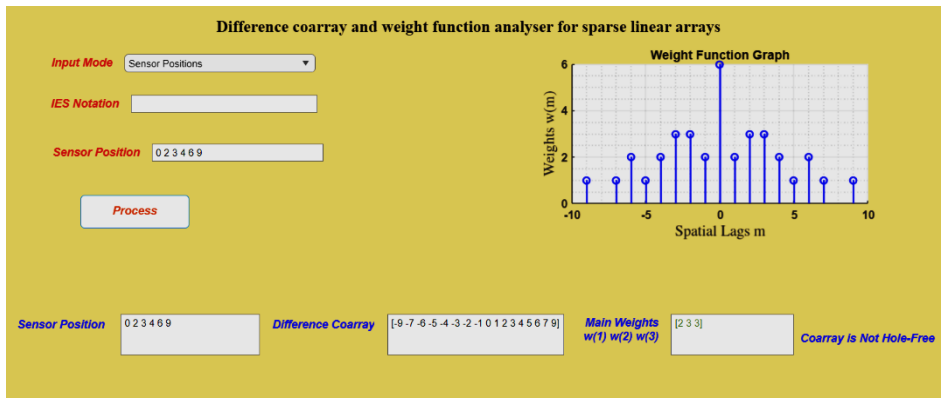

Fig. 6. Coprime array with $N = 6$ obtained using co-primes (2, 3)

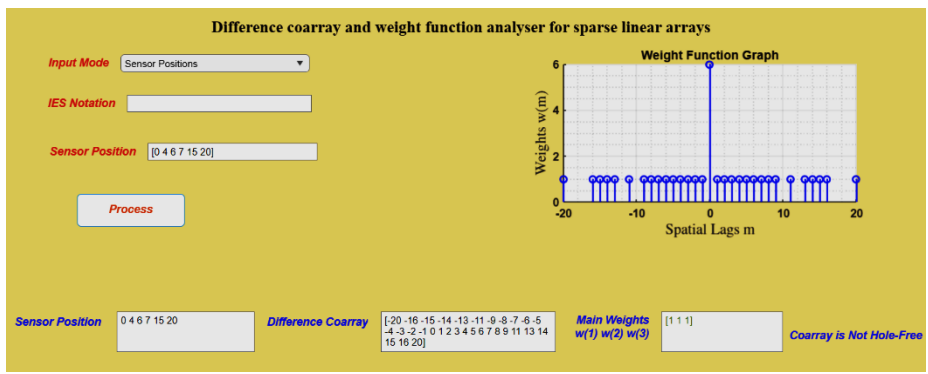

Fig. 7. Optimal 6-element ODNRA

WCSAs keep mutual coupling at bay due to their peculiar design [14]. One such array, namely, $z_6$ can provide $O(N)$ DOFs. The $z_6$ array with ten elements ($N = 10$) has sensors at [-7, -4, 0, 5, 10, 15, 20, 25, 28, 31]. The sensor positions are in line with the expected aperture of $5N - 12$. Care must be taken while analyzing such arrays as there is a chance to misread the array aperture to be 31, whereas in fact it is 38. because the first sensor is located at -7. Fig. 8 shows the coarray analysis of the 10-element $z_6$ array considered above. The aperture, hole locations, and primary weights exactly match the entries in the last row of Table II of Kulkarni and Vaidyanathan [14], as expected.

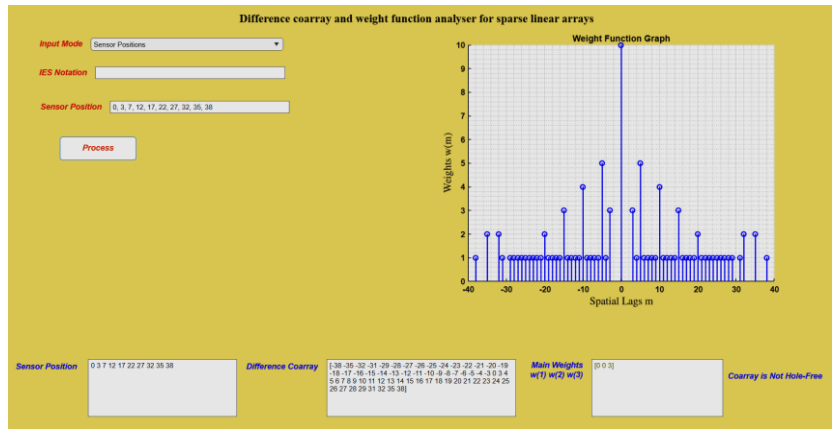

Fig. 8. 10-element $z_6$ by Kulkarni and Vaidyanathan

C. *Entries using the IES notation*

The simulator also takes input in the form of IES notations. For example, a 15-element ULA with sensors positions [0, 1, 2, …, 14] can be represented in the IES format as $\{1^{14}\}$. This array can be inputted to the GUI by selecting 'IES notation' from the drop down menu and entering the string 'ones(1, 14)'. Fig. 9 shows the coarray parameters.

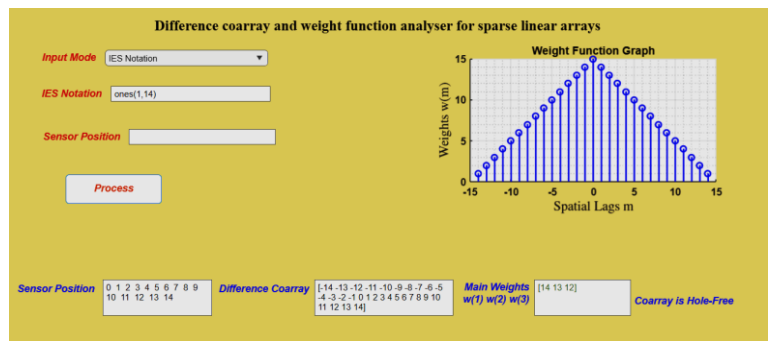

Fig. 9. 15-element ULA in IES format

Similarly, the array with sensors at [0, 2, 4, 6, 8, 10, 12, 14] can be represented in the IES format as $\{2^7\}$. The GUI input string would be '2*ones(1,7)' and the output as in Fig. 10. Note that this array offers the same aperture as the 15-element ULA (Fig. 9). However, since it's sensors are placed at alternate grid positions, it is regarded as a sparse array. This array has no adjacent sensors (sensors with unit spacing), implying $w(1) = 0$. Additionally, there are no sensor pairs with a separation distance of three units, and hence, $w(3) = 0$. Consequently, this sparse array is far less susceptible to mutual coupling than the 15-element ULA. It is well-known in sparse array literature that the first three weights (primary weights) empirically determine the array's susceptibility to mutual coupling [13], [16], [17]. Lower values of these primary weights correspond to higher immunity against the effects of mutual coupling.

Although the array [0, 2, 4, 6, 8, 10, 12, 14] provides the same aperture as the ULA and has significantly lower mutual coupling, it's DCA consists of holes at all odd spatial lags, namely, $\pm 1, \pm 3, \pm 5, \pm 7, \pm 9, \pm 11$, and $\pm 13$, as shown in Fig. 10. While this array is not suited for coarray-based DOA estimation, a simple trick can render its DCA to be hole-free. Adding an extra sensor either at position {1} or {13} can generate all the missing spatial lags. Fig. 11 shows the coarray properties of the array [0, 1, 2, 4, 6, 8, 10, 12, 14], inputted using the IES format $\{1, 1, 2^6\}$. The extra sensor at {1} covers all previously missing spatial lags and gives a hole-free DCA. However, its inclusion in the array increases the number of closely-spaced sensor pairs; leading to higher primary weights. Nevertheless, these primary weights are far lower than those of the 15-element ULA shown in Fig. 9. It can be concluded that the GUI presented here is versatile and accurate.

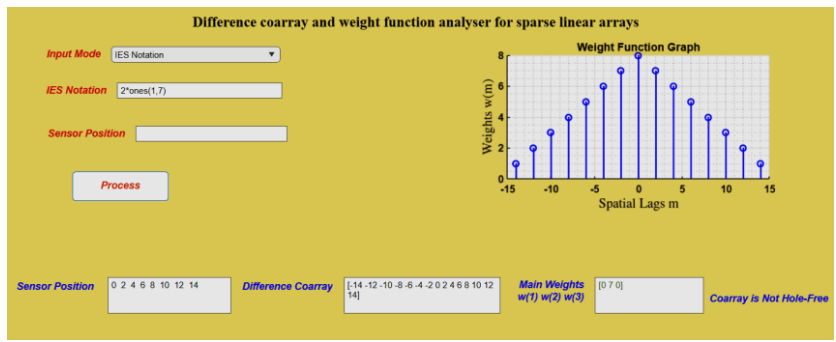

Fig. 10. Coarray analysis of the sparse array [0, 2, 4, 6, 8, 10, 12, 14]

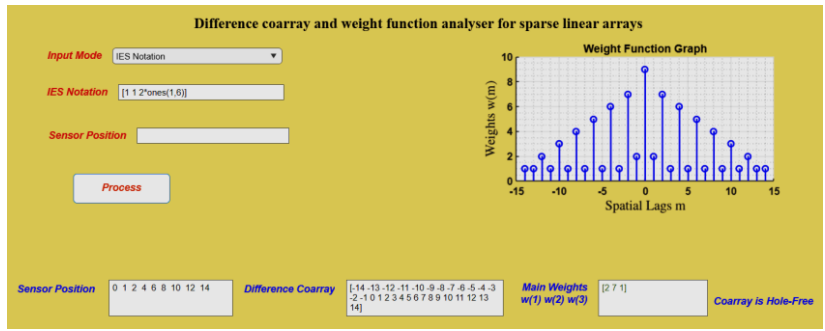

Fig. 11. Hole-Free DCA due to the inclusion of the extra sensor at {1}

Many other sparse arrays such as the maximum interelement spacing criterion (MISC) array [16], the enhanced MISC (xMISC) array [38], the generalized enhanced MISC (GEMISC$_E$) array [39], the two-fold redundant array [40], the ternary redundant array [41] etc., were analyzed using the simulator and the results were consistent with the theoretical understanding. All these examples validate the functional correctness of the simulator. The tool is user friendly and computationally light as it produces the output almost instantly. This makes it ideal for users to play around with the input array configurations and study how the choice of sensor positions influence coarray properties.

## V. LIMITATIONS AND FUTURE SCOPE

While the tool is currently complete in itself, there is always a room for further refinement. For instance, a function to calculate the coarray redundancy of the AUT or a function to compute the coupling leakage factor ($\mathcal{L}$) of the AUT could be added. Additionally, the scope of the simulator could be extended to include sum co-arrays (SCAs), sum difference co-arrays (SDCAs), and sparse MIMO radar arrays. Additional functions such as the provision to perform DOA estimation using the AUT for a specific user-defined scenario (source angles and number of sources) could be added. Similarly, simultaneous input of two arrays and side-by-side comparison of their respective coarray properties can be incorporated into the tool.

## VI. CONCLUSION

This paper explains the design and validation of a MATLAB-based GUI for sparse array analysis in the coarray domain. The simulator was tested against various test inputs before packaging it as a MATLAB app. The final GUI app and its various functions have been tested thoroughly taking known arrays as an input and validating whether the simulator output matches prior knowledge about that array. Additionally, a case study has been presented where the proposed tool can also be used for sparse array design through human-in-the-loop techniques and visual feedback. The developed tool is highly advantageous for researchers and students. It can be an essential resource for teaching sparse array analysis to the novice. In the future, similar tools can be developed in open source languages such as Julia or Octave so that the tools remain accessible to everyone.

## APPENDIX A.1

```matlab
%% Program to find sensor positions from IES
clc; clear all; close all;
a = [ones(1,17) 3*ones(1,18) 2*ones(1,2) ones(1,2)]; %Array configuration in IES form
n = length(a)+1; % number of sensors is n
b = zeros(1,n);% actual positions of physical sensors. All initialized to zero
for i=2:n
    b(i)=b(i-1)+a(i-1);
end
disp('The physical sensors are at:');
b % display the actual sensor positions
```

## APPENDIX A.2

```matlab
clc; clear all; close all;
% N=input('Enter the number of sensors');
a = [0 1 2 4 8 14 18 20 21 22] ; %Declare the physical array.
N = numel(a); % N denotes the number of sensors in the array
x = a - a.'; % This commands generates a N*N matrix.
d = reshape(x,[1 N*N]); % This gives the difference set as a row vector of size 1*N^2.
```

```
dca = unique(sort(d)); %This gives the DCA i.e., sorted and non-repeating entries.
```

APPENDIX A.3

```
%%% program to find and plot the weight function graph
w = histc(d,dca);
stem(dca,w,'b','LineWidth',1.5);
xlabel('Spatial lags $m$','FontSize',14,'Interpreter','Latex')
ylabel('Weights $w(m)$','FontSize',14,'Interpreter','Latex')
set(gca,'Xtick',[-40:5:40]);
ylim([0 numel(a)])
xlim([-max(a) max(a)])
grid on
grid minor
```

APPENDIX A.4

```
if(length(dca)==(2*max(a)+1))
    disp('Coarray is hole-free')
else
    disp('Coarray has holes')
end
```

APPENDIX A.5

```
y = zeros(1,max(a)+1);
y(a+1)=1;
w = round(xcorr(y));
main_weights = [w(max(a)+2)  w(max(a)+3)  w(max(a)+4)];
```

**REFERENCES**

[1] R. A. Monzingo, R. L. Haupt, and T. W. Miller, *Introduction to Adaptive Arrays*. IET Digital Library, 2011. doi: 10.1049/SBEW046E, 10.1049/SBEW046E.
[2] F. Gross, *Smart Antennas for Wireless Communications: With MATLAB*. McGraw Hill Professional, 2005.
[3] S. Haykin, J. P. Reilly, V. Kezys, and E. Vertatschitsch, 'Some aspects of array signal processing', *IEE Proc. F - Radar Signal Process.*, vol. 139, no. 1, pp. 1–26, Feb. 1992, doi: 10.1049/ip-f-2.1992.0001.
[4] L. C. Godara, 'Smart Antennas', CRC Press. Accessed: Mar. 31, 2016. [Online]. Available: https://www.crcpress.com/Smart-Antennas/Godara/9780849312069
[5] C.-L. Liu and P. P. Vaidyanathan, 'Cramér–Rao bounds for coprime and other sparse arrays, which find more sources than sensors', *Digit. Signal Process.*, vol. 61, pp. 43–61, Feb. 2017, doi: 10.1016/j.dsp.2016.04.011.
[6] P. Raiguru, D. C. Panda, and R. K. Mishra, 'Multi-Source Detection Performance of Some Linear Sparse Arrays', *IETE J. Res.*, vol. 0, no. 0, pp. 1–12, Mar. 2022, doi: 10.1080/03772063.2022.2038701.
[7] P. Raiguru et al., 'Hole-Free DCA for Augmented Co-Prime Array', *Circuits Syst. Signal Process.*, vol. 41, no. 5, pp. 2977–2987, May 2022, doi: 10.1007/s00034-021-01909-0.
[8] P. Pal and P. P. Vaidyanathan, 'Coprime sampling and the music algorithm', in *2011 Digital Signal Processing and Signal Processing Education Meeting (DSP/SPE)*, Jan. 2011, pp. 289–294. doi: 10.1109/DSP-SPE.2011.5739227.
[9] P. P. Vaidyanathan and P. Pal, 'Sparse Sensing With Co-Prime Samplers and Arrays', *IEEE Trans. Signal Process.*, vol. 59, no. 2, pp. 573–586, Feb. 2011, doi: 10.1109/TSP.2010.2089682.
[10] Piya Pal, P. Pal, P.P. Vaidyanathan, and P. P. Vaidyanathan, 'Nested Arrays: A Novel Approach to Array Processing With Enhanced Degrees of Freedom', *IEEE Trans. Signal Process.*, vol. 58, no. 8, Art. no. 8, Aug. 2010, doi: 10.1109/tsp.2010.2049264.
[11] M. Yang, L. Sun, X. Yuan, and B. Chen, 'Improved nested array with hole-free DCA and more degrees of freedom', *Electron. Lett.*, vol. 52, no. 25, pp. 2068–2070, 2016, doi: 10.1049/el.2016.3197.
[12] Z. Peng, Y. Ding, S. Ren, H. Wu, and W. Wang, 'Coprime Nested Arrays for DOA Estimation: Exploiting the Nesting Property of Coprime Array', *IEEE Signal Process. Lett.*, vol. 29, pp. 444–448, 2022, doi: 10.1109/LSP.2021.3139577.
[13] S. Wandale and K. Ichige, 'A Generalized Extended Nested Array Design via Maximum Inter-Element Spacing Criterion', *IEEE Signal Process. Lett.*, vol. 30, pp. 31–35, 2023, doi: 10.1109/LSP.2023.3238912.
[14] P. Kulkarni and P. P. Vaidyanathan, 'Weight-Constrained Sparse Arrays For Direction of Arrival Estimation Under High Mutual Coupling', *IEEE Trans. Signal Process.*, vol. 72, pp. 4444–4462, 2024, doi: 10.1109/TSP.2024.3461720.
[15] A. Patwari and P. Kunchala, 'Novel Sparse Linear Array Based on a New Suboptimal Number Sequence with a Hole-free Difference Co-array', *Prog. Electromagn. Res. Lett.*, vol. 116, pp. 23–30, 2024, doi: 10.2528/pierl23102706.

[16] Z. Zheng, W.-Q. Wang, Y. Kong, and Y. D. Zhang, 'MISC Array: A New Sparse Array Design Achieving Increased Degrees of Freedom and Reduced Mutual Coupling Effect', *IEEE Trans. Signal Process.*, vol. 67, no. 7, pp. 1728–1741, Apr. 2019, doi: 10.1109/TSP.2019.2897954.
[17] C. Liu and P. P. Vaidyanathan, 'Super Nested Arrays: Linear Sparse Arrays With Reduced Mutual Coupling—Part I: Fundamentals', *IEEE Trans. Signal Process.*, vol. 64, no. 15, pp. 3997–4012, Aug. 2016, doi: 10.1109/TSP.2016.2558159.
[18] R. Cohen and Y. C. Eldar, 'Sparse Array Design via Fractal Geometries', *IEEE Trans. Signal Process.*, vol. 68, pp. 4797–4812, 2020, doi: 10.1109/TSP.2020.3016772.
[19] A. Patwari, 'Sparse Linear Antenna Arrays: A Review', in *Antenna Systems*, IntechOpen, 2021. doi: 10.5772/intechopen.99444.
[20] M. G. Amin, *Sparse Arrays for Radar, Sonar, and Communications*. John Wiley & Sons, 2023.
[21] M. Pesavento, M. Trinh-Hoang, and M. Viberg, 'Three More Decades in Array Signal Processing Research: An optimization and structure exploitation perspective', *IEEE Signal Process. Mag.*, vol. 40, no. 4, pp. 92–106, Jun. 2023, doi: 10.1109/MSP.2023.3255558.
[22] A. M. Elbir, K. V. Mishra, S. A. Vorobyov, and R. W. Heath, 'Twenty-Five Years of Advances in Beamforming: From convex and nonconvex optimization to learning techniques', *IEEE Signal Process. Mag.*, vol. 40, no. 4, pp. 118–131, Jun. 2023, doi: 10.1109/MSP.2023.3262366.
[23] W. Liu, M. Haardt, M. S. Greco, C. F. Mecklenbräuker, and P. Willett, 'Twenty-Five Years of Sensor Array and Multichannel Signal Processing: A review of progress to date and potential research directions', *IEEE Signal Process. Mag.*, vol. 40, no. 4, pp. 80–91, Jun. 2023, doi: 10.1109/MSP.2023.3258060.
[24] R. Rajamaki and V. Koivunen, 'Symmetric Sparse Linear Array for Active Imaging', Jul. 2018, Accessed: May 02, 2019. [Online]. Available: /documents/symmetric-sparse-linear-array-active-imaging
[25] R. Rajamäki and V. Koivunen, 'Sparse Symmetric Linear Arrays With Low Redundancy and a Contiguous Sum Co-Array', *IEEE Trans. Signal Process.*, vol. 69, pp. 1697–1712, 2021, doi: 10.1109/TSP.2021.3057982.
[26] N. Mohsen, A. Hawbani, X. Wang, M. Agrawal, and L. Zhao, 'Optimized Sparse Nested Arrays for DoA Estimation of Non-circular Signals', *Signal Process.*, vol. 204, p. 108819, Mar. 2023, doi: 10.1016/j.sigpro.2022.108819.
[27] P. Gupta and M. Agrawal, 'Design And Analysis of the Sparse Array for DoA Estimation of Noncircular Signals', *IEEE Trans. Signal Process.*, vol. 67, no. 2, pp. 460–473, Jan. 2019, doi: 10.1109/TSP.2018.2883035.
[28] M. Yang, L. Sun, X. Yuan, and B. Chen, 'A New Nested MIMO Array With Increased Degrees of Freedom and Hole-Free Difference Coarray', *IEEE Signal Process. Lett.*, vol. 25, no. 1, pp. 40–44, Jan. 2018, doi: 10.1109/LSP.2017.2766294.
[29] A. Patwari and R. R. Gudheti, 'Novel MRA-Based Sparse MIMO and SIMO Antenna Arrays for Automotive Radar Applications', *Prog. Electromagn. Res.*, vol. 86, pp. 103–119, 2020, doi: 10.2528/PIERB19121602.
[30] M. Ebrahimi, M. Modarres-Hashemi, and E. Yazdian, 'Optimal placement of sensors to enhance degrees of freedom in monostatic collocated MIMO radar', *Digit. Signal Process.*, vol. 142, p. 104224, Oct. 2023, doi: 10.1016/j.dsp.2023.104224.
[31] 'Array Factor'. Accessed: Jan. 09, 2025. [Online]. Available: https://in.mathworks.com/matlabcentral/fileexchange/38332-array-factor
[32] *Barkhausen-Institut/AntennaArraySimulator*. (Sep. 30, 2024). Python. Barkhausen Institut gGmbH. Accessed: Jan. 09, 2025. [Online]. Available: https://github.com/Barkhausen-Institut/AntennaArraySimulator
[33] 'GitHub - zinka/arraytool: Python based package for phased array antenna design and analysis'. Accessed: Jan. 09, 2025. [Online]. Available: https://github.com/zinka/arraytool/tree/master
[34] M. T. Yalcinkaya, P. Sen, M. Ramzan, and G. P. Fettweiss, 'Highly Portable Open Source Array & Phased Antenna Simulator', in *IEEE EUROCON 2023 - 20th International Conference on Smart Technologies*, Jul. 2023, pp. 450–454. doi: 10.1109/EUROCON56442.2023.10198953.
[35] A. Harrison, *RadarBook/software*. (Jan. 07, 2025). Jupyter Notebook. Accessed: Jan. 09, 2025. [Online]. Available: https://github.com/RadarBook/software
[36] *arrayfactor: Array Factor Calculator*. Python. [Microsoft :: Windows]. Available: https://pypi.org/project/arrayfactor/
[37] M. Ebrahimi and M. Karimi, 'Optimally Dense Nonredundant Sparse Sensor Array Designs', *IEEE Sens. J.*, vol. 24, no. 18, pp. 28952–28959, Sep. 2024, doi: 10.1109/JSEN.2024.3431272.
[38] S. Wandale and K. Ichige, 'xMISC: Improved Sparse Linear Array via Maximum Inter-Element Spacing Concept', *IEEE Signal Process. Lett.*, vol. 30, pp. 1327–1331, 2023, doi: 10.1109/LSP.2023.3316018.
[39] P. Zhao, Q. Wu, G. Hu, L. Wang, and L. Wan, 'Generalized Hole-Free Sparse Antenna Array Design With Even/Odd Maximum Interelement Spacing', *IEEE Antennas Wirel. Propag. Lett.*, vol. 24, no. 3, pp. 716–720, Mar. 2025, doi: 10.1109/LAWP.2024.3514150.
[40] D. Zhu, S. Wang, and G. Li, 'Multiple-Fold Redundancy Arrays With Robust Difference Coarrays: Fundamental and Analytical Design Method', *IEEE Trans. Antennas Propag.*, vol. 69, no. 9, pp. 5570–5584, Sep. 2021, doi: 10.1109/TAP.2021.3083819.
[41] S. HOU, X. ZHAO, Z. LIU, Y. PENG, and N. WANG, 'Ternary redundant sparse linear array design with high robustness', *J. Northwest. Polytech. Univ.*, vol. 41, pp. 125–135, Feb. 2023, doi: 10.1051/jnwpu/20234110125.